\documentclass[10pt]{IEEEtran}
\usepackage{array}
\usepackage[switch,mathlines]{lineno}
\usepackage{soul}
\usepackage{graphicx}
\usepackage{cite}
\usepackage[cmex10]{amsmath}
\usepackage{algorithm}

\usepackage{algpseudocode}
\usepackage{fixltx2e}
\usepackage{CJK}
\usepackage[cmex10]{amsmath}
\usepackage{bm}
\usepackage{color}
\usepackage{amsmath,amsthm,amssymb,amsfonts}

\usepackage{arydshln} %to include dashed lines in arrays

\usepackage{soul}
\usepackage{tikz}
\usepackage{booktabs}% http://ctan.org/pkg/booktabs
\usepackage{tabularx}% http://ctan.org/pkg/tabularx
\usepackage{tikz}

\usepackage{float}
\usepackage{multirow}
\usepackage{placeins}

\usepackage{makecell}
\usepackage{changepage}

\begin{document}

%Jaber Valinejad, Student member, IEEE, Lamine Mili, Life Felow, IEEE

\title{\huge{Research Needed in Computational Social Science for Power System Reliability, Resilience, and Restoration }}
%Social-Constrained Optimal Power Flow 

\author{Jaber Valinejad, Student member, IEEE, Lamine Mili, Life Fellow, IEEE,  C. Natalie van der Wal
%\vspace{-1.1cm}

\thanks{\hspace{-8pt}\underline{~~~~~~~~~~~~~~~~~~~~~~~~~~~~~~~~~~~~~~~~~~~~~~~~~~~}}% <-this % stops a space
%\thanks{}
\thanks{The work presented in this paper was funded by the National Science Foundation (NSF) under Grant No. 1917308.

J. Valinejad and L. Mili are with the Bradley Department of Electrical and Computer Engineering, Virginia Tech, Northern Virginia Center, Greater Washington D.C., VA 22043, USA. J.  Valinejad  is also  with  the Department  of  Computer Science at the same university.(email:{JaberValinejad,lmili}@vt.edu).

Natalie van der Wal is with Delft University of Technology, Faculty of Technology, Policy and Management, dept. Multi-Actor Systems, Netherlands.(e-mail: C.N.vanderWal@tudelft.nl)}
%C. Huang, X. Chen, M. Korkali, C. Tong and L. Min are with
%Lawrence Livermore National Laboratory, Livermore, CA 94550 USA (e-mail:
%{huang38,chen73,korkali1,tong10,min2}@llnl.gov)
%Credence. US-Ireland Research and Development Partnership Program (centre to centre), funded by Science Foundation Ireland (SFI) and The National Science Foundation (NSF) under the grant number 16/US-C2C/3290.

}

\markboth{
%IEEE Transactions on Power Systems,~Vol.~XX, No.~Y, MONTH ZZZ 2020
}%
{Shell \MakeLowercase{\textit{\textit{et al.}}}: Bare Demo of IEEEtran.cls for Journals}
\maketitle

\begin{abstract}
In the literature, smart grids are modeled as cyber-physical power systems without considering the computational social aspects. However, end-users are playing a key role in their operation and response to disturbances via demand response and distributed energy resources. Therefore, due to the critical role of active and passive end-users and the intermittency of renewable energy, smart grids must be planned and operated by considering the computational social aspects in addition to the technical aspects. The level of cooperation, flexibility, and other social features of the various stakeholders, including consumers, prosumers, and microgrids, affect the system efficiency, reliability, and resilience. In this paper, we design an artificial society simulating the interaction between power systems and the social communities that they serve via agent-based modeling inspired by Barsade's theory on the emotional spread. The simulation results show a decline in the consumers' and prosumers' satisfaction levels induced by a shortage of electricity. It also shows the effects of social diffusion via the Internet and mass media on the satisfaction level. In view of the importance of computational social science for power system applications and the limited number of publications devoted to it, we provide a list of research topics that need to be achieved to enhance the reliability and resilience of power systems' operation and planning. 
%analyze a case study to demonstrate the proposed approach. 

\end{abstract}
%\vspace{-0.2cm}
%\vspace{-0.9cm}

\section{Introduction}
%\vspace{-0.2cm}
A power grid is a cyber-physical-social system in its essence. Indeed, humans contribute to all the processes involved in electric generation, transmission, distribution, and consumption, from planning to operation and maintenance \cite{8464032}. Therefore, considering the social aspects is essential for both the generation side and the end-user side. On the generation side, prosumers' (active end-users) social behavior affects the power system control and operation in real-time via demand responses and batteries of electric vehicles. The cooperation and participation of  end-users, whether passive or active, can contribute to their active involvement in the system ancillary services, such as voltage and frequency stability \cite{suryanarayanan2016cyber}. Putting the consumers (passive end-user) and the prosumers at the center of a power system study is also necessary for assessing its reliability, resilience and associated community resilience \cite{jaber2020cybernetic}. The social aspects account for the involvement, not only of the active end-users, but also of the primary energy industries, e.g., the coal industry and other organizations connected to the power system \cite{xue2017beyond,zhang2017social}.  Indeed, both the end-users and the primary energy industry shape the smart grid operation's objectives significantly. In addition to these stakeholders, human errors in the power industry influence  maintenance, emergency dispatch operation, and rolling blackout. In addition, human errors may contribute to cascading failures leading to blackouts. Therefore, by considering the social aspects, power systems can be made more efficient, reliable, resilient, and sustainable.\par
Emotions are the foundation for psychological and social behaviors of the consumers and prosumers  \cite{silver2002nationwide,Braden2015,8744387}.  The power systems end-users' social behavior and emotional status affect its operation in various ways. For example, different types of emotions determine the end-users' satisfaction level \cite{jaber2020,rand2014positive,Fredrickson2002}. Based on neuroscience, psychology, and social science, the consumers' satisfaction level and social behaviors, influences each other via the use of mass media platforms. Therefore, there is a need to consider the social spread of social aspects in a power system. In this paper, we extend the study described in our previous papers regarding these social aspects \cite{jaber2020}. Specifically, we designed an artificial society simulating the interrelationship between power systems and social communities via agent-based modeling. Our model is inspired by Barsade's theory on emotional spread. We aim to show the effect of computational social science with this example. Simulation results show that the average level of satisfaction of the end-users decline with the increase of the load shedding and the degree of accessibility of the mass media platforms. We then provide a list of research topics that need to be achieved to enhance the reliability and resilience of the operation and planning of power systems. This research is needed due to the limited amount of publications in this area. \par
The remainder of this paper is structured as follows. Section II provides a simple model and the results for the satisfaction of end-users considering social diffusion and electricity availability. This aims to show the effect of computational social science on the power system through a simple case study. Section III provides the need to take account of the computational social science in power systems and future research.  The conclusions are provided in Section IV. 
 %\vspace{-0.3cm}

\section{Building the artificially society to model the Satisfaction of the End-Users}
This section investigates the effect of computational social science on the power system and vice versa. 
To incorporate computations social science and collective behavior in power systems operation and planning, we propose to use generative computational social science \cite{epstein1996growing}. In this approach, we create an artificial society from the bottom-up - from the literature or starting at the theory. 
%- as opposed to creating a model top-down from data.
The artificial society can be used for virtual experiments via agent-based modeling and simulation \cite{8226990}. Generative computational social science is an appropriate and promising method widely accepted by sociology, complex systems, emergence, and evolutionary programming \cite{7491286,van2017simulating}. By creating an artificial society, we are able to model the micro-macro levels of social-power systems, look for emergent patterns, and investigate the aggregated effects \cite{epstein1996growing}. We apply this approach to the well-being and resilience of power systems end-users. We propose to base the social behavior of the whole social system on Barsade's theory of emotion spread \cite{Barsade1998}.  We believe that in the bulk power system, the collective behavior of consumers and prosumers is of great significance.\par
In the literature, some researchers advocate the use of an artificial society to model social systems' collective behavior, the interaction between agents, and human response. For instance, Gao and Liu \cite{paperasli} propose a network-based model for people's collective behavior during a disaster, which encompasses features like emotion, risk perception, and information-seeking behavior. Bosse and colleagues $et$ $al.$ \cite{natalie_13} propose an multi-agent-based model considering fear, intention, and information behavior as the main characteristics of people. Jiang and Jiang \cite{6869018} initiate a multi-agent diffusion model to simulate collective behavior and contagion among a set of social actors. \par

The social features $S_{t}$ of the stakeholders are modeled as  
%\vspace{0.2cm}

%\footnotesize
%\noindent
\begin{eqnarray}
 S_{n}(t+\Delta t)=[\varpi_{1}f(C_{n}(t),P_{n}(t),S_{n}(t)) \nonumber  \\ 
 +\varpi_{2}g(C_{m}(t),S_{m}(t)) - S_{n}(t)]\Delta t+ S_{n}(t) ,
 \label{eq:1} 
\end{eqnarray}
%\normalsize
where $C_{n}$ denotes the cyber variables and parameters - such as the availability of the Internet and the promotion of fake news in the mass media-, $P_{n}$ denotes the physical variables and parameters - such as the availability of electricity -, $S_{tn}$ denotes the social variables and parameters - such as cooperation and emotion - for agent n at time t. $f(C_{n}(t),P_{n}(t),S_{n}(t))$ is a combination of the cyber, physical, and social variables. Note that the subscript n denotes the agent for which this is calculated, and m designates  all the other agents that influence the agent n. As for the second term in Eq. (1), $g(C_{m},S_{m})$ represents the effect of the social feature S of all the other agents m on the agent n and $\varpi_{1}$ and $\varpi_{3}$ are weighting factors. All cyber, physical, and social variables have values within the interval [0 1]. Note that the function $f(C_{n}(t),P_{n}(t),S_{n}(t))$ can be highly nonlinear, which can make the optimization problem difficult to solve. {This general model is based on a vast literature related to neuroscience, psychological, social science, and discussions with colleagues in these areas \cite{van2017simulating,natalie_13}. This model has a strong foundation on various psychological and social neuroscience theories or hypotheses, i.e., Fredrickson’s  broaden-and-build theory, Barsade emotional contagion theory, and Damasio’s somatic marker hypothesis. To make our model biologically plausible, we consider social diffusion where the social behaviors are propagated by mirror neurons in the brain discovered by neuroscientists \cite{cacioppo2005social,harmon2007social}.}  \par
{To account for consumers' and prosumers' levels of dissatisfaction, we leverage the bottom-up approach that is derived from Barsade's and Fredrickson's theories on emotion spread \cite{Barsade1998, Fredrickson2002}. The community's collective dissatisfaction level depends on the homogeneity and heterogeneity of each agent's emotional state and mood and their minimum, maximum, and mean level. Fredrikson's theory indicates that the interest of a person  to something or someone leads to the motivation to explore, enhancing her physical, social, intellectual, and psychological skills. As a result, positive emotions lead to an increase in personal resources. It is a belief that positive emotions broaden thought-action proceedings, attention, and cognition, both at present and in the future. That is the case with negative emotion, i.e., the dissatisfaction level of consumers and prosumers.} Let D denote the dynamical change of the dissatisfaction level, which takes higher values when the person is more dissatisfied. It is expressed as

%\footnotesize
\begin{equation}
D_{(t+1)n}=\frac{g(S_{tm})}{\varpi_{2}}(\hat{D}_{tn}-D_{tn}) \Delta t +D_{tn} , 
 \label{eq:1} 
\end{equation}
%\normalsize

%\vspace{-0.2cm}
where t is the time. Here, $\hat{D}_{tn}$ indicates the amount of the effect  of emotion diffusion and electricity on the consumers and prosumers, which is given by 

%\footnotesize
\begin{equation}
% \begin{multlined}
%\begin{align*}
  \begin{array}{l}
\hat{D}_{tn}=\underbrace{\varpi_{1}(1-E_{tn})}_{f(C_{tn},P_{tn},S_{tn})}+\underbrace{\varpi_{2}(\frac{\sum_{m} \gamma_{tnm} D_{tm}}{\sum_{m} \gamma_{tnm}})}_{g(S_{tm})},
\label{eq:4} 
%\end{multlined}
%\end{align*}
\end{array}
\end{equation}
%\normalsize

where $w_{1}$, and $w_{2}$ are the weighting factors. Note that the level of dissatisfaction is contingent on the physical-social dependence,i.e., the electricity availability ($E_{tn}$). In addition to the internal cyber-physical-social dependence, $\hat{D}_{tn}$ is influenced by the diffusion of the social-cyber dependence, including the diffusion of dissatisfaction  of the other agents ($g(C_{tm},S_{tm})$). $\gamma_{tnm}$ is the contagion strength from  agent m to n.   $g(C_{tm},S_{tm})$ denotes the emotion spread, which is the weighted dissatisfaction of each agent based on the bottom-up approach \cite{natalie_13}. {Emotions in the group can be expressed, received, or transferred in such a way as to  affect the energy level of the group.} The dissatisfaction is propagated through the mass media platforms via the Internet if it is available. In summary, people can communicate through the mass media channels and express their emotions, which in turn affects the social dissemination. Our computational model is verified by Case Study 1 that is taken from \cite{natalie_13}.  

\begin{figure}[h]
\centering
\includegraphics[width= 0.8 \columnwidth]{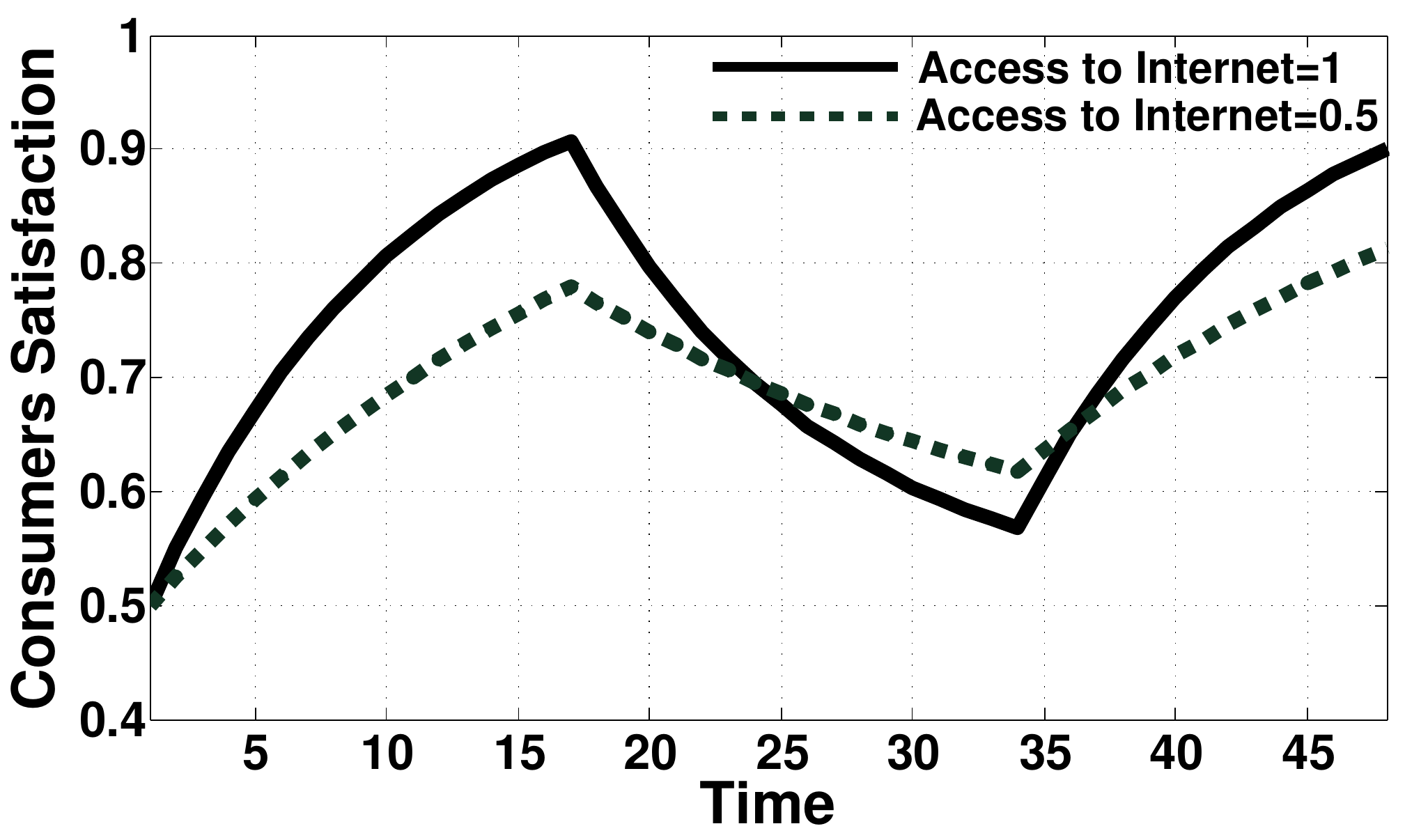}
\vspace{0.5cm}
\setlength{\abovecaptionskip}{-15pt}
\caption{The average  level  of end-user's satisfaction   during  48 hours for two different scenarios, i.e., when the Internet is fully accessible and when it is not. The electricity is not available between hours 14 and 34. }
%\vspace{-0.7cm}
\label{fig:Fig_34a1}
\end{figure}

\textit{Description and Analysis of  Case Study}: customer’s satisfaction level over 48 hours with load shedding. We consider the following scenario: 9 end-users - grouped as 1-3, 4-6, and 7-9 -  are living in different residential areas. Each area’s end-users are in contact with each other via mass media platforms. Note that the level of satisfaction and load shedding stay within the interval [0 1]. The initial customers’ satisfaction level for the service provided by the utilities is set as 0.5, reflecting a ‘neutral’ level, not high, not low. The level of load shedding between hours 17 and 34 has increased from 0 to 0.5 while that between the hours 1 and 17 and between 34 and 48 is equal to 0. Figure~\ref{fig:Fig_34a1} displays two graphs of the average level of the customer’s satisfaction for 48 hours; either for customers that have full or limited access to the Internet. In the latter case, the level of access is fixed to 0.5. The bold curve shows that from hours 1 to 17, the average level of consumer’s satisfaction calculated using the model given by (2) increases to 1. After that, it decreases until hour 34 because of an increase in the level of load shedding. Then, it increases after hour 34 to reach 90\% at hour 48. As for the dotted graph, we notice that the emotion spead reduces, which in turn decreases the intensity of the diffusion impact of the availability of electricity on the satisfaction level of consumers. According to the argument “opposites being beneficial”, emotional heterogeneity may contribute to the emotional group balance, the appropriate group outcomes, and the process \cite{o1998group}.
\par  

%\vspace{-0.3cm}
%\vspace{-0.1cm}

\section{Future Research: the Need of Computational Social Science Methods in Power Systems}
\begin{figure}[h]
\centering
\includegraphics[width= 0.75 \columnwidth]{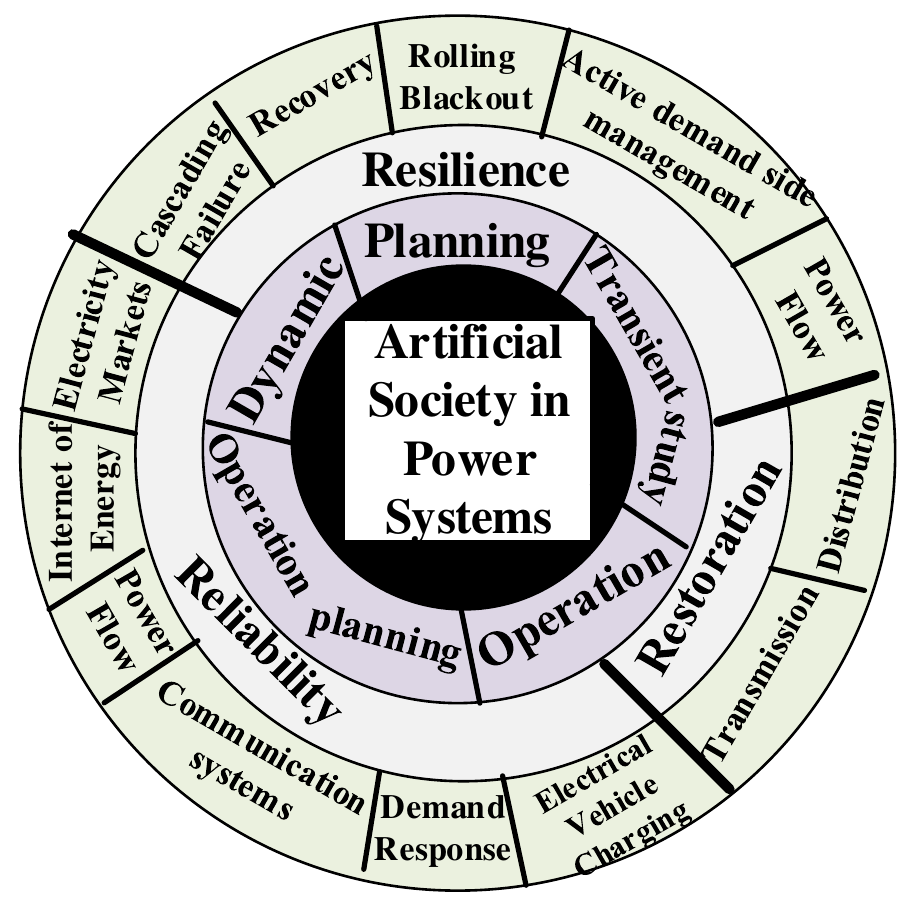}
\vspace{0.5cm}
\setlength{\abovecaptionskip}{-15pt}
\caption{An overview of the research topics needed in computational social science for power system reliability, resilience, and restoration modeling using artificial society methodology.   }
\label{fig:Fig_34}
\end{figure}

%\vspace{-0.2cm}
In the previous section, we modeled the level of social diffusion and dissatisfaction of  end-users to load shedding to investigate their effects on power systems. In this section, we propose the research topics needed in computational social science  for power system reliability, resilience, and restoration; see Figure~\ref{fig:Fig_34}. Nowadays, the exchange of information via the Internet between producers and consumers is gradually increasing so that the power industry is turning to an Industrial Internet of Things industry.Because of the close interaction of producers and end-users in various applications of power systems, there is an inevitable demand to incorporate  computational social science in power system operation and planning analysis. Artificial society should be incorporated into power system analysis, planning, and control with different time scales as seen in the first circular layer of Figure~\ref{fig:Fig_34}. In addition, various operating and planning results are caused by the use of artificial society in the study of resilience, reliability, and restoration of the power system, as shown in the second circular layer. Resilience covers cascading failure, recovery, rolling blackout, and active demand-side management, to name a few.  On the other hand, reliability consists of demand response, electrical vehicle charging, investment planning, communications systems, Internet of energy, electricity markets, to name a few. Finally, computational social science should be modeled in the field of  both distribution and transmission restoration.  For each research topic, we explain why we need social science, psychological behavior of stakeholders, and, therefore, an artificial society.

\textit{Socio-Technical Power Flow}: The availability of electricity affects both the mental well-being and physical well-being of consumers, prosumers, and the community as a whole \cite{liddell2015living,critchley2007living,ahlborg2015provision,Scofield2018,ortiz2017review,7900389}. When a disaster strikes, typically, a certain number of generating units and transmission lines are disconnected, inducing the shedding of generation and load. Evidently, this  shedding should be achieved by minimizing its impact on the social well-being and the community resilience subject to power flow constraints, as shown in \cite{jaber2020cybernetic}. This is achieved by optimizing the social well-being and the community resilience subject to power system constraints. 

\textit{Investment in Microgrids and Distributed Energy Resources}: The investment in microgrids and distributed energy resources play a key role in the social, economic, and infrastructure resilience. Technical and economic aspects of this investment have been considered in the literature \cite{dagoumas2019review}. In addition to these aspects, the social well-being and community resilience should be considered in the investment in microgrids as well.\par
%,valinejad2017generation,babatunde2019comprehensive
\textit{Power System Planning}: Similar to the investment in microgrids,  power system planners  should also consider enhancing community resilience and social welfare as two main objectives when planning the investment in new generating units, transmission lines, and charging stations for electric vehicles. Investment in power system expansion should be performed fairly in this regard.\par 

\textit{Rolling Blackout and Load Shedding}: 

Rolling blackouts affect the power system end-users. To reduce the community vulnerability and increase the resilience, rolling blackouts should be carried out fairly.  The levels of cooperation \cite{7782466,nowak2006five,parkash2015cooperation,Braden2015,fehr1999theory,axelrod1981evolution} and flexibility \cite{Braden2015,reiter2003predicting,duarte2018flexibility} of the end-users  can help the utilities to enhance both the infrastructure and the community resilience. These levels are influenced by the levels of satisfaction \cite{Barsade1998,silver2002nationwide,8876861,xu2019crowd}, trust \cite{glaeser2000measuring,hardin2002trust,luhmann2018trust}, and learning 
\cite{Braden2015,hoffmann2017learn,8663430,hoffmann2015tale} of the end-users, to name a few. \par

\textit{Demand response}: {The study of the demand response in the literature is limited to the economic \cite{ye2016game,mcpherson1993evaluating}and sustainability \cite{lokeshgupta2018multi} aspects . Most of the demand response models ignore the social science aspect despite its great importance \cite{jordehi2019optimisation} .} Participation in demand response scheduling depends directly on the level of flexibility, cooperation, empathy\cite{banfield2012role,hojat2016empathy,peters2014compassion,rochat2002various,goleman2006emotional} , emotional status, and habits of the community. Hence, there is a need to consider them in the demand response modeling, which involves the achievement of trade-offs between various objective functions, i.e., social well-being, sustainability, and cost.

\par

\textit{Transactive Energy}: Another application of computational social science in power systems is transactive energy, where stakeholders such as wholesale and retail sellers, prosumers, and buyers of energy services are involved. The modeling of the decision-making of the latter relies heavily on the modeling of their social behavior \cite{marzband2018framework,yang2020blockchain}. \par

\textit{Electricity Markets}: Power market, similar to any other market, should consider the psychological aspects of the investors, such as trust and satisfaction \cite{litvine2011helping,hobman2016uptake}. In turn, the culture of each community can influence these psychological aspects \cite{furaiji2012empirical,ng2003satisfying}. The winner of this electric market is the stakeholder who takes the consumers and their social behavior into account.  \par 

\textit{Electrified Transportation System with Large Penetration of Electric Vehicles}: Long charging time may make an electric vehicle's owner anxious. Indeed, the limited distance to drive per each charge can increase the owner's concern \cite{seebauer2018psychology}. By driving until discharging the electric vehicle battery following the vehicle-to-grid program, the owner may face problems to reach a destination. Furthermore, when using a blockchain for electric vehicles, the owners may be anxious about their privacy and information \cite{knirsch2018privacy}. This anxiety and concern regarding the charging, discharging, and privacy affect the owner's actions and decision to participate in the vehicle to grid program \cite{axsen2013social}. The participation of the owner also depends on the level of cooperation, flexibility, and trust to the aggregator.\par

\textit{Renewable Energy}: The electric generation of renewable energy units, such as wind turbine and photovoltaic units, experience random intermittences. This makes their productions highly uncertain and difficult to predict \cite{hu2020bayesian,xu2018propagating}. Consequently, the frequency of the power system may experience a large variability. An effective way to address this problem is to enhance the active demand side management to frequency changes via better cooperation between the consumers. \par

\textit{Pandemic Planning}: An infectious outbreak such as a  pandemic influenza impacts the primary energy and, in turn, the electric supply chain. Consequently, it affects the electric generation and decreases the resiliency of the power system. In fact, pandemic makes a disconnection between the fuel supply chain (the primary energy) and the electric sector (the secondary energy) \cite{kelley2014pandemic}.  {Hence, the normal operation of the bulk power system is disturbed. This is the reason why studying the impact of the pandemic on electric power systems is emphasized by selected Federal agencies such as Health and Human Services (HHS) \cite{HHS}, Department of Homeland Security (DHS) \cite{DHS}, Department of Energy (DOE) \cite{DOE}, Department of Transportation (DOT) \cite{DOT}, and Department of the Interior(DOI) \cite{DOI}.}\par

\textit{Voltage and Frequency Stability}: An epidemic, e.g., COVID-19, induces a decrease in industrial and commercial loads and a possible increase in residential loads. It may produce an increase in the harmonic voltage levels at specific frequencies. This in turn increases the vulnerability of the power system to voltage and frequency instabilities. The cooperation and flexibility of the end-users who are willing to engage in the demand response program is essential to enhance power system stability margins in real-time.\par

\textit{Cascading Failures}: Cascading failures in power systems, e.g., the blackout that the USA and Italy experienced in 2003, can induce failures in other critical infrastructures such as transportation, water supply, health care systems, financial services, and communication systems \cite{dobson2005loading,chen2013composite,nedic2006criticality}. The utilities and end-users' role is to take appropriate actions to minimize the economic losses that may result, which is dependent on human activity and the mental states, habit, and culture. For example, if a blackout occurs, microgrids can provide electricity to critical loads such as hospitals, gas stations, police stations, data centers, to name a few \cite{chen2013assessing}. Overloading is one of the triggers for a cascading failure in power systems, leading to equipment outages and blackouts. It turns out that active demand side management can mitigate this risk. Furthermore,  the detrimental consequences of electric power outages on a community can be alleviated via enhancing the level of cooperation, flexibility, and empathy. If power outages occur, we must put critical loads as a priority to supply to enhance community resilience.  The prosumers within the community can share their electricity with vulnerable people, e.g., elderly and  handicapped and sick people, as a priority. Besides, the consumers within the community can reduce their loads.

\textit{Heating, Ventilation, and Air Conditioning (HVAC) System}: HVAC systems should be designed by considering the comfort \cite{zhang2010comfort}, work performance \cite{magraner2010comparison}, satisfaction, mental and physical well-being of its customers. \par 

\textit{Economic Dispatch and Unit Commitment}:  Active end-users can supply some parts of the electricity to the load as part of the grid's real-time operation. Hence, the prosumers' high level of cooperation and flexibility can increase the economic dispatch and unit commitment efficiency if it is properly modeled.

\textit{Environmental Protective Policies and Regulations}: End-users may promote investment in energy hub resources to improve environmental conservation programs.

\textit{Recovery to a Specific Disaster}: A line, a tower, a transformer, a substation have different times to repair, typically from small to large \cite{coffrin2011strategic,gomez2002voltage,ahn2007recovery}. To estimate these times to repair, there is a need for social neuroscience model to account for the past experience and the level of cooperation and flexibility and the level of emotion of the utility workforce, as well as the level of information \cite{ryan2013information,cho2018dynamics,timbus2009active} and resources \cite{kavousi2018stochastic,tyagi2013sludge} available to them. Pre-event prevention and mitigation of the impact of the hurricane on the power system and the emergency services are event modifiers between the event and post-event time period.  We propose to leverage an artificial society to model workforce behavior that influences power system vulnerability.

\textit{Active Demand Side Management}:
Frequency regulation is of high importance to power system stability. Active consumers in smart grids can play a significant role in providing ancillary services for power system transient stability \cite{xu2010demand}. Although  frequency regulation is important in normal power system operation, it is more crucial and challenging during a disaster. The consumers who receive signals by smart meters can decrease the risk of electricity outage by participating in an active demand side management \cite{di2018active}. They can turn off unnecessary electric devices for a few hours as soon as they receive a signal about the disaster from the emergency services or utilities. The participation of consumers in active demand side management to provide ancillary services for frequency regulation is entwined with their social behavior.

\textit{Restoration}: Restoration of a power system following a blackout induced by various extreme events is a complex process. It consists of phases, i.e., planning to restart and reintegration of the bulk power system, retaining critical sources of power (degraded level), and restoration after stabilizing at some degraded level \cite{317561}. The level of experience, knowledge, learning rate, the cooperation of the workforce, experts, and operators influence the quality of restoration. 

The Internet of Things, big data and cloud computing, complex networks, blockchain, instant data and edge computing, artificial intelligence, and power system SCADA are cyber-physical systems. However, a power system is a cyber-physical-social system where the social system refers to stakeholders' social behaviors, which affect the way the power system operates. Obviously, these stakeholders' social behavior has a significant impact on the efficiency, reliability, and resilience of a power system \cite{mili2018integrating,mili2011taxonomy,valinejad2020stochastic,hamborg2020rethinking,guidotti2019integration}. We extend these features to power systems stakeholders. We classify the social elements in different manners. The classification of social traits can be target-based, time-based, or stakeholder-based. In the target-based classification, social features are categorized into resiliency-based, sustainability-based, economic-based, efficiency-based, stability margin-based, and reliability-based characteristics. In the time-based classification, social components are categorized into planning-based, operational planning-based, real-time operation-based, and real-time, dynamics, and transients-based characteristics. In the stakeholder-based classification, social features are  categorized into end-users-based, primary energy provider-based, secondary energy provider-based, other organizations –based behaviors. Table~\ref{tab:aaa} shows the stakeholder-based classification of the social features. Each of these stakeholders, humans, organizations, and societies, are  defined by their specific social characteristics.

\begin{table*}[h!]
%    \centering
\caption{The stakeholder-based classification of the social features for power system operation and planning  }
 \begin{adjustwidth}{-0.1cm}{}
\label{tab:aaa}
    \resizebox{0.8\textwidth}{!}{\begin{minipage}{\textwidth}
    \begin{tabular}{|l|l|l|l|}
    \hline
    \hline
        \makecell{Stakeholders} & \makecell{Sub-Stakeholders}&\makecell{Social Features} &\makecell{References} \\ \hline \hline
        End users & \makecell{Prosumers, Consumers, \\Electrical Vehicle Owners \\(Active, passive End users), DERs} & \makecell{Willingness, Convenient, Emotion, Fear, Satisfaction , Empathy, Risk Perception, Price-Information \\-seeking, Social Wale fare, Comfort (m), Collaboration, Experience, Motivation, Learning,\\ Flexibility , Bias, Cooperation , Belief, Mass media diffusion, Mental Health, Physical Health} &\makecell{\cite{Braden2015,xu2019crowd,lebow2005reason},\\\cite{kjell2013exploring,paperasli,7491286,van2017simulating},\\\cite{natalie_13,cacioppo2005social,harmon2007social}
        } \\ \hline
        \makecell{Secondary\\ Energy \\Providers} & \makecell{Retailer, Utilities, Genco, Transco, Disco, MG, \\load serving entities, Aggregator, ISOs \\Regional transmission organizations (RTO) }& \makecell{Price-Information-seeking, Energy-Information-seeking, Reliability, Confidence,\\ Learning, Cooperation (co), Privacy, Work efficiency, Risk Perception, Emotion,\\ Fear, Collaboration, Experience, Mental Health, Physical Health} &\makecell{\cite{silver2002nationwide,rand2014positive,levine2018signaling},\\ \cite{southward2017assessing,Barsade1998}  }  \\ \hline
        \makecell{Primary\\ Energy\\ Companies} & \makecell{Coal industry, Natural Gas industry,\\ Hydro, Biofuels and waste, Oil }& \makecell{Price-Information-seeking, Fuel-Information-seeking, Reliability,\\ Confidence, Learning, Cooperation, Work efficiency, Risk Perception,\\ Emotion, Collaboration, Experience, Mental Health, Physical Health}&\makecell{\cite{xue2017beyond,jaber2020cybernetic,kelley2014pandemic}, \\ \cite{epstein1996growing,6869018} }  \\ \hline
        Other Entities & \makecell{City councils , Edge Computing \\Service
        provider (ESP)} &\makecell{ Emission-Information-seeking, Reliability, Confidence, Learning, Cooperation,\\ Privacy, Work efficiency, Risk Perception, Emotion, Flexibility,  Collaboration, Experience}&\makecell{ \cite{ring1992structuring,fischer2012risk,paperasli}}  \\ \hline     
    \end{tabular}
    \end{minipage}}
%\vspace{-0.7cm}
\end{adjustwidth}
\end{table*}
\normalsize

The social behavior of the end-users is individual-based. When they are highly  satisfied,  the end-users, such as prosumers and electric vehicle owners, increase their level of flexibility, cooperation, and trust to secondary energy providers such as electric utilities or retailers \cite{rand2014positive,lebow2005reason,levine2018signaling,kjell2013exploring,southward2017assessing}. Consequently, they are more willing to participate in demand response, active demand side management, and vehicle to grid programs, to name a few. Furthermore, if the consumers and the prosumers exhibit a high level of empathy and collaboration,  they may be willing to enhance the power system efficiency, reliability, and resiliency. \par
On the other hand, how the end-users think affects their decision-making related to the power system operation \cite{dadkhah2017network,ma2019analysing,ding2019multi}. Their decision-making is also affected by their experience and their learning behavior. In addition, the secondary energy providers can increase the motivation of the end-users to contribute to better efficiency of the demand response. During a disaster, when the end-users experience fear and a shortage of electricity, they perceive risk \cite{slovic1982study,renn1998role,paperasli}. This risk, in turn, increases their level of cooperation \cite{ring1992structuring,fischer2012risk}. Hence, they are more prone to participate in an active demand side management program and share electricity with their neighbors who have experienced an outage. In addition to the quality of service, social diffusion through the mass media platforms affects their satisfaction level \cite{paperasli}. Social diffusion means that an end-user's emotion affects the feeling of others about power system services. More detailed discussions are provided in \cite{jaber2020}. The social behavior of secondary energy providers, primary energy companies, and other entities are organizational-based. The social behaviors of each stakeholder can affect its service quality and decisions that are very important to the performance, reliability, and resilience of power systems, as shown in Table~\ref{tab:aaa}.  
%\vspace{-0.4cm}

\section{Conclusions}
%\vspace{-0.15cm}

In his paper, we highlight the need to account for the cyber-physical-social dependence in power engineering in order to model and optimize the efficiency, resilience, sustainability, reliability,  stability, and economic aspect of the power infrastructure and the associated social community. To meet that need, we build an artificial society based on the bottom-up approach to model the end-users' dissatisfaction through a simple case study to investigate the  interdependence between computational social models and power systems. We provide a comprehensive list of research topics needed in computational social science for various power system operation and planning activities.  We believe that our approach represents a paradigm shift in that it integrates power systems and artificial society science into social computing methods. Each of the applications stated needs a significant research effort to mature.  Our simulation results based on the agent interactions show emerging trends, i.e., collective behaviors that are not predictable from the  individual agent rules. Integrating an artificial society into the power system planning and operation process allows us to derive new hypotheses and test different scenarios that can be costly and difficult to carry out in the real world. In the next section, we elaborate on how computational social science impacts to power systems in various applications.
%\vspace{-0.6cm}

\addcontentsline{toc}{section}{References}
%\nocite{*}
%\printbibliography
%\bibliographystyle{plain}
\bibliographystyle{IEEEtran}
\bibliography{neighbourhood} 

\end{document}